\documentstyle[prb,aps]{revtex}

\begin{document}
\draft
\title{Aging in a Two-Dimensional Ising Model with Dipolar Interactions} 

\author{Julio H. Toloza, Francisco A. Tamarit, Sergio A. Cannas\cite{auth2}}  

\address{Facultad de Matem\'atica, Astronom\'\i a y F\'\i sica, 
Universidad Nacional de C\'ordoba, Ciudad Universitaria, 5000 
C\'ordoba, Argentina}
\date{\today}
\maketitle

\begin{abstract} 
Aging in a two-dimensional Ising spin model with both ferromagnetic exchange 
and antiferromagnetic dipolar interactions is established and investigated 
via Monte Carlo simulations. The behaviour of the autocorrelation function 
$C(t,t_w)$ is analyzed for different values of the temperature, the waiting 
time $t_w$ and the quotient $\delta=J_0/J_d$, $J_0$ and $J_d$ being the 
strength of exchange and dipolar interactions respectively. Different  
behaviours are encountered for $C(t,t_w)$ at low temperatures as $\delta$ is 
varied. Our results show that, depending on the value of $\delta$, the 
dynamics of this {\bf non-disordered} model is consistent either with a slow 
domain dynamics characteristic of ferromagnets or 
with an activated scenario, like that proposed for spin glasses. 

\end{abstract}

\pacs{PACS numbers: 75.40.Gb, 75.40.Mg, 75.10.Hk}

Microscopic long ranged interactions are always of interest in different 
fields of physics because they can give rise to a variety of unusual 
macroscopic behaviours. Perhaps the better example  in condensed matter are 
dipole-dipole interactions. In particular, the competition between long-range 
antiferromagnetic dipolar interactions and short-range ferromagnetic exchange 
interactions can give rise to several interesting magnetic phenomena. Recent 
works in two dimensional uniaxial spin systems, where the spins are oriented 
perpendicular to the lattice and coupled with these kind of interactions, 
have shown a very rich phenomenological scenario concerning both its 
equilibrium statistical mechanics\cite{Kashuba,MacIsaac} and non-equilibrium 
dynamical properties\cite{Sampaio}. Magnetization processes in these kind of 
systems are of interest due to aspects related to information storage in 
ultrathin ferromagnetic films. Moreover, there are several contexts in which 
a short-ranged tendency to order is perturbated by a long-range frustrating 
interaction. Among others, model systems of this type has been proposed to 
study avoided phase transitions in supercooled liquids\cite{Kivelson} and 
charge density waves in doped antiferromagnets\cite{Chayes,Zachar,Pryadko}

The above mentioned systems can be described by an Ising like Hamiltonian of 
the type 
\begin{equation}
H = - J_0\sum_{<i,j>}{\sigma_i\sigma_j} + 
J_d\sum_{(i,j)}{\frac{\sigma_i\sigma_j}{r_{ij}^3}}
\end{equation}
where the spin variable $\sigma_i=\pm 1$ is located at the site $i$ 
of a square lattice, the sum $\sum_{<i,j>}$ runs over all pairs of nearest 
neighbor sites and the sum $\sum_{(i,j)}$ runs over all distinct pair of 
sites of the lattice; $r_{ij}$ is the distance (in crystal units) between 
sites $i$ and $j$; $J_0>0$ and $J_d>0$ are the ferromagnetic exchange and 
antiferromagnetic dipolar coupling parameters respectively. For simplicity, 
we rewrite this Hamiltonian as follows:
\begin{equation}
\label{hamilton}
H = - \delta\sum_{<i,j>}{\sigma_i\sigma_j} + 
\sum_{(i,j)}{\frac{\sigma_i\sigma_j}{r_{ij}^3}}
\end{equation}
with $\delta=J_0/J_d$. There are few numerical results concerning the 
equilibrium statistical mechanics, {\it i.e.} the finite temperature phase 
diagram, of this model. In a recent work MacIsaac and coauthors\cite{MacIsaac}
have shown that the ground state  of Hamiltonian (\ref{hamilton}) is the 
antiferromagnetic state for $\delta<0.85$. For $\delta>0.85$ the 
antiferromagnetic state becomes unstable with respect to the formation of 
striped domain structures, that is, to state configurations with spins 
aligned along a particular axis forming a ferromagnetic strip of constant 
width $h$, so that spins in adjacent  strips are anti aligned,  forming a 
super lattice in the direction perpendicular to the strips. They also showed
that striped states of increasingly higher thickness $h$ becomes more stable 
as $\delta$ increases from $\delta=0.85$. Moreover, they showed that the 
striped states are also more stable than the ferromagnetic one for arbitrary 
large values of $\delta$, suggesting such a phase to be the ground state of 
the model for $\delta>0.85$. Monte Carlo calculations on finite lattices at 
low temperature\cite{MacIsaac,Sampaio} gave further support to this proposal, 
at least for intermediate values of $\delta$. Furthermore, such simulations 
have shown that  striped phases of increasingly higher values of $h$ may 
become thermodynamically stable at {\bf finite} temperatures for intermediate 
values of $\delta$. This results are in agreement with other analytic 
ones\cite{Kashuba,Chayes}. For low values of $\delta$ the system presents an 
antiferromagnetic phase at low temperatures. At high temperatures, of course, 
the system becomes paramagnetic.

The dynamics of the model is characterized by the formation and growth of 
magnetic domains, due to the competition between the exchange and the dipolar
interactions, which at low temperatures generate very large relaxation times. 
In a Monte Carlo study of the time evolution of the magnetization, Sampaio 
{\it et al}\cite{Sampaio} have shown the existence of two different type of 
relaxation, according to the value of $\delta$. For $\delta>\delta_c\sim 2.7$ 
the magnetization relaxes exponentially, with a  relaxation time that depends 
both on the temperature and $\delta$. For $\delta<\delta_c$ the magnetization 
presents a power law decay, with an exponent independent of $\delta$. They 
also showed the existence of strong hysteresis effects in the presence of an 
external magnetic field for $\delta>\delta_c$. Since the relaxation time in 
this case increases with $\delta$,  the area of the hysteresis loops increases
with $\delta$. Hysteresis can also be detected in a simulation by looking at
the evolution of the system starting from different initial configurations. In
Fig.1 we show snapshots of the evolution of a $N=64\times 64$ system and
$\delta=4$. For $\delta<\delta_c$ hysteresis effects becomes very small and the
magnetization presents a quasi linear dependence on the magnetic field, up to
the saturation field (see Fig.2).  Hysteresis is a typical non-equilibrium
phenomenon associated to domain dynamics with large relaxation times. In this
scenario one also expects the presence of {\bf aging} effects, that is,
history-dependence in the time evolution of the response functions after the
system has been quenched into some non-equilibrium state.

Aging phenomena has been vastly studied in disordered systems such as spin
glasses (see Ref.\cite{Cugliandolo} and references therein), which are
essentially out of equilibrium on experimental time scales. However, they
appear also in the phase ordering kinetics of {\bf ordered} systems, such as
the  Ising {\bf ferromagnet}\cite{Newman,Blundell,Bray}, associated with a slow
domain dynamics. Aging can be observed in real systems through different
experiments. A typical example is the zero-field-cooling\cite{Lundgren}
experiment, in which the sample is cooled in zero field to a sub-critical
temperature at time $t_0$. After a waiting time $t_w$ a small constant magnetic
field is applied and subsequently the time evolution of the magnetization is
recorded. It is then observed that the longer the waiting time $t_w$ the slower
the relaxation.

Although aging can be detected through several time-dependent quantities, a
straightforward way to establish it in a numerical simulation  is to calculate
the spin autocorrelation function
\begin{equation}
\label{autocor}
C(t,t_w) = \frac{1}{N}\sum_i\langle\sigma_i(t+t_w)\sigma_i(t_w)\rangle
\end{equation}
where $< \cdots >$ means an  an average over different realizations of the
thermal noise and $t_w$ is the  waiting time,  measured from some  quenching
time $t_0=0$.

In this work we present the results of  Monte Carlo simulations in the
two-dimensional Ising spin model defined by the Hamiltonian (\ref{hamilton}) on
a $N=20\times 20$ square lattice with free boundary conditions. We chose the
heat-bath algorithm for the spin dynamics and time is measured in Monte Carlo
steps per site. The quantity (\ref{autocor}) is averaged over 100 samples; for
each run the system is initialized in a random initial configuration
corresponding to a quenching from infinite temperature to the temperature $T$
at which the simulation is done. We analyze the behaviour of $C(t,t_w)$ as a
function of the observation time $t$, for different values of $t_w$, $\delta$
and $T$.

At enough high temperatures we find that the system does not present aging,
that is, for any value of $\delta$ there is a temperature above which
$C(t,t_w)$ is independent of $t_w$, as expected in a paramagnetic phase. At low
temperatures we find different types of aging behaviours as $\delta$ is varied.

The typical behaviours of $C(t,t_w)$ are illustrated in Figures 3 and 4, for
$T=0.5$ and different values of $\delta$ (waiting times $t_w=5^n
(n=2,\ldots,6)$). We also analyzed, for the same values of $T$ and $\delta$,
the time evolution of the magnetization per site $m(t)$  and the staggered
magnetization per site $m_s(t)$ starting from different initial conditions , in
order to characterize the different relaxation regimes.

In Fig.3a we see the typical behaviour of $C(t,t_w)$ in the antiferromagnetic
state. The characteristic signature of $C(t,t_w)$ in this regime is the
appearance of a plateau at some intermediate value of $t$ independent of $t_w$
($t\sim 10^3$ for $\delta=-1$), where $C(t,t_w)$ remains constant for a period
of time that depends on $T$ and $\delta$; after such period $C(t,t_w)$ relaxes
to zero. We also see a dependency on $t_w$, that is, aging. The inset shows the
evolution of $m(t)$ (filled circles) and $m_s(t)$ (open circles) starting from
a random initial condition. We also analyzed the evolution of the same
quantities starting from a fully magnetized initial state ($m(0)=1$). We did
not find any hysteresis effect in this region, the only difference being a much
more slower convergence of $m_s(t)$ towards a constant value. These kinds of
behaviours appear for negative values  and also for small positive values of
$\delta$ at enough small temperatures, in agreement with previous calculations
of the phase diagram of the model\cite{MacIsaac}. The existence of these
plateaus suggests some type of quasiequilibrium state. This behaviour is rather
unusual in this kind of system and it is probably related to the long range
character of the interactions.

In Fig.3b we see the typical behaviour of $C(t,t_w)$ in the paramagnetic state,
where it does not depend on $t_w$. In this case both $m(t)$ and $m_s(t)$
converge fast to zero, independent of the initial configuration, as expected.
This result also illustrates the characteristic shape of the low temperature
phase diagram of the  model (see Fig.6 of Ref.\cite{MacIsaac}): following a
line of constant temperature in the $(\delta,T)$ diagram one can go from the
antiferromagnetic region for low values of $\delta$ to the striped one for high
values of $\delta$, passing through a {\bf paramagnetic} region at some
intermediate values of $\delta$. For low enough temperatures it is expected a
transition line between the antiferromagnetic and the striped regions, without
an intermediate paramagnetic one. It is worth to note the {\bf logarithmic}
decay of $C(t,t_w)$. This is a characteristic feature of this dynamics, that
is, the relaxation of $C(t,t_w)$ is always slow, even in the paramagnetic
region.

In Fig.4 we show $C(t,t_w)$ for different values of $\delta$ corresponding to
the striped region, below\cite{Sampaio} (figures 4a and 4b) and above (figures
4c and 4d) $\delta_c\sim 2.7$. The behaviour of $m_s(t)$ in all these cases
shows no traces of antiferromagnetic ordering. On the other hand, the behaviour
of $m(t)$ starting from a random ($m(0)=0$) and a fully magnetized ($m(0)=1$)
states shows a clear differentiation between both regions: while for
$\delta>\delta_c$ strong hysteresis effects appear (see also Fig.1), for
$\delta<\delta_c$ such effects are negligible, in agreement with the results of
Ref.\cite{Sampaio} (see also Fig.2).

We found that $C(t,t_w)$ obeys a different type of dynamic scaling law for
every one of the two striped regions, as can be seen from the data collapse of
figure 5. For $\delta<\delta_c$ $C(t,t_w)$ obeys the dynamic scaling
\begin{equation}
C(t,t_w) \propto c_{\delta} \left( \log{(t)}/\log{(\tau(t_w))} \right),
\end{equation}
(see figures 5a and 5b), while for $\delta>\delta_c$ it obeys the following
scaling
\begin{equation}
C(t,t_w) \propto c_{\delta} \left( t/\tau(t_w) \right),
\end{equation}
(see figures 5c and 5d), where the time scale $\tau(t_w)$\cite{Rieger} is shown
in the insets.  We observed, for values of $\delta$ far enough of $\delta_c$,
the scaling form $\tau(t_w) \sim t_w^a$. The log-log linear fitting from Fig.5
gives the following values: $a=0.83 \pm 0.02$ for $\delta=2.0$; $a=0.93\pm
0.05$ for $\delta = 3.5$ and $a=1.00\pm 0.05$ for $\delta=4.0$. This values
implies for $\delta$ well below $\delta_c$ a logarithmic scaling $C(t,t_w)
\propto c \left( \log{(t)}/\log{(t_w)}\right)$ (Fig.5a). The scaling form of
$C(t,t_w)$ deviates from the previous one as $\delta$ approaches $\delta_c$
(Fig.5b). For $\delta>\delta_c$ we observe a scaling form 
$C(t,t_w)\propto c\left( t/t_w^a\right)$ (figures 5c and 5d), with a exponent
$a$ that approaches the unity as $\delta$ increases.

It has been proposed\cite{Fisher,Bray} that aging phenomena are based on a slow
domain growth at low temperatures, where after certain time $t$ a
characteristic domain size $L(t)$ is reached. The time evolution of quantities
like the autocorrelation function will then present a crossover from dynamical
processes characterized by length scales smaller than the already achieved
domain size to processes at larger scales dominated by domain growth through
the movement of domain walls. In this scenario, scaling arguments\cite{Bray}
leads to the following expected dependency of $C(t,t_w)$:
\begin{equation}
C(t,t_w) \propto c \left( L(t)/L(t_w)  \right),
\end{equation}
at least for large values of $t$ and $t_w$.

Our results for $\delta>\delta_c$ are consistent with a $t/t_w$ scaling for
large values of $\delta$, where the ferromagnetic short range interactions are
dominant. This type of scaling, which is associated with an algebraic growth of
the domain size $L(t)\propto t^\psi$,  appears in the slow domain growth
dynamics of disordered systems with a ferromagnetic ground
state\cite{Newman}-\cite{Bray}.

Much more interesting is the logarithmic scaling form $\log{(t)}/\log{(t_w )}$
when $\delta<\delta_c$. This scaling is associated with a logarithmic time
dependence of the domain size $L(t)\propto(\log{t_w})^\psi$ predicted by an
activated scenario\cite{Fisher} in spin glasses, in which disorder and
frustration generate active droplets excitations with a broad energy
distribution.

In summary we have shown that the interplay between short- and long-range
competitive interactions in an ordered system give rise to different kinds of
aging at low temperatures. The different behaviours appear to be related to
different domain dynamics as the relative strengths of the interactions are
changed. In particular, for intermediate values of $\delta$, where the
strengths are comparable, the competition between them generate a slow
relaxation dynamic consistent with an activated scenario, like that proposed
for spin glasses. This result suggests a possible relationship between the
microscopic dynamic properties of this model  and that of disordered frustrated
systems. In this sense, it would be of interest to investigate, for instance,
the possible existence of broad energy distributions of low-lying excitations
in the present model. Works along these directions are in progress and would be
published elsewhere.

Fruitful suggestions from Leticia F. Cugliandolo and Daniel Stariolo are
aknowledged. This work was partially supported by grants from Consejo Nacional
de Investigaciones Cient\'\i ficas y T\'ecnicas CONICET (Argentina), Consejo
Provincial de Investigaciones Cient\'\i ficas y Tecnol\'ogicas (C\'ordoba,
Argentina) and  Secretar\'\i a de Ciencia y Tecnolog\'\i a de la Universidad
Nacional de C\'ordoba (Argentina).

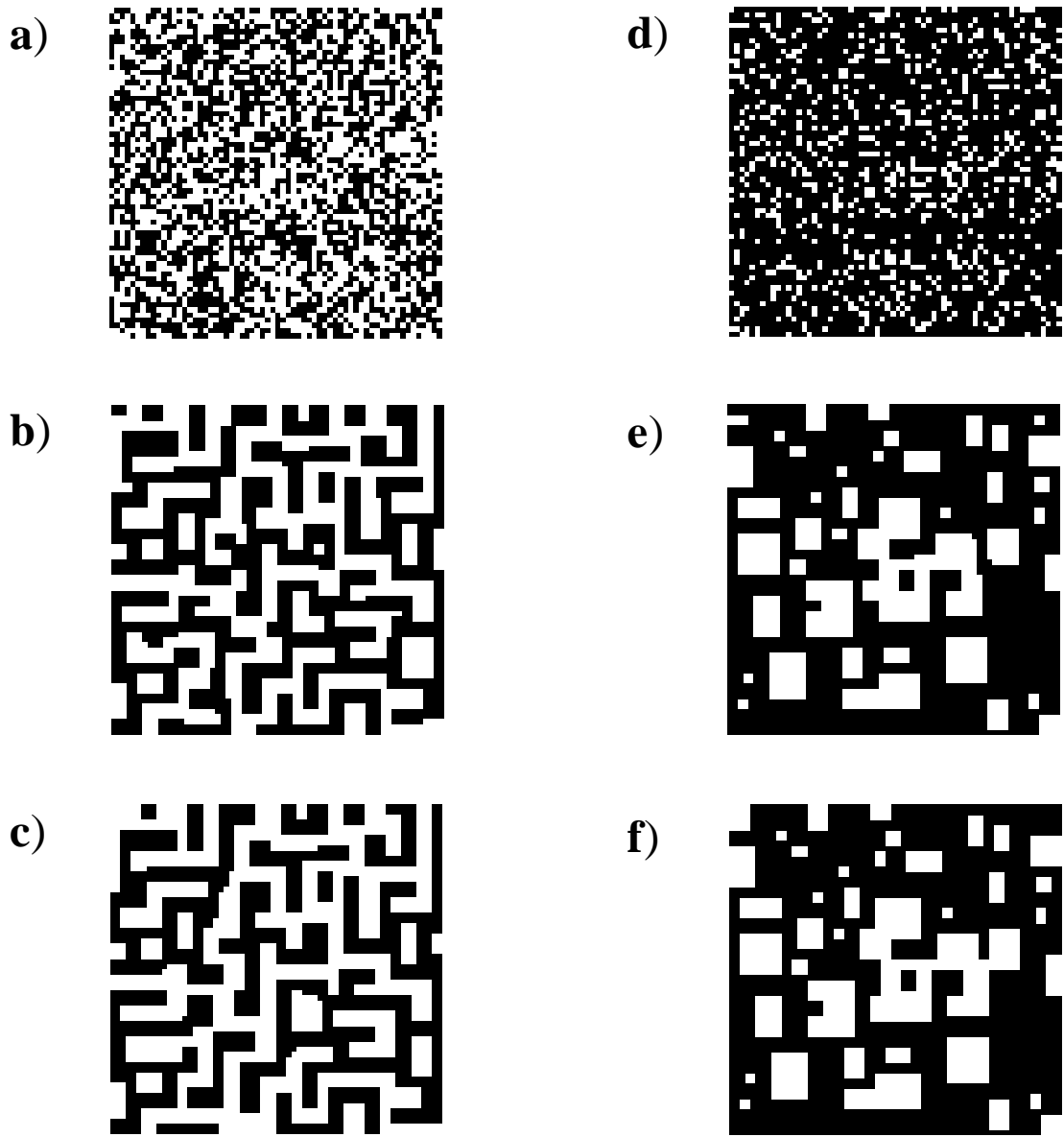
\begin{figure}
\caption{Snapshots of the time evolution of a $64\times 64$ system with 
 $\delta=4$, $T=0.5$ for different initial conditions. Figures (a), (b) and (c)
 correspond to an initial configuration with mean magnetization per site 
 $m(0)=0$, for $t=1, 10$ and $1000$ Monte Carlo steps per site respectively. 
 Figures (d), (e) and (f) correspond to an initial configuration with 
 $m(0)=0.6$ and the same values of $t$.}
\label{fig1}
\end{figure}

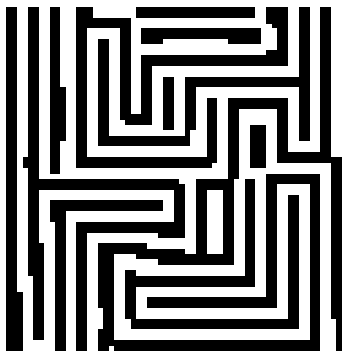
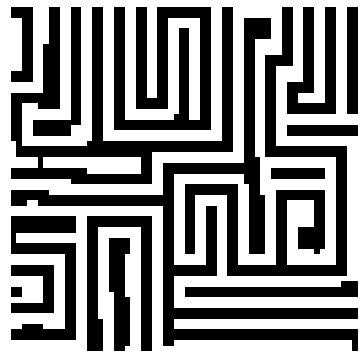
\begin{figure}
\caption{Snapshots of the time evolution of a $64\times 64$ system with 
 $\delta=2$, $T=0.5$ for different initial conditions. Figures (a), (b) and (c)
 correspond to an initial configuration with mean magnetization per site 
 $m(0)=0$, for $t=1, 10$ and $1000$ Monte Carlo steps per site respectively. 
 Figures (d), (e) and (f) correspond to an initial configuration with 
 $m(0)=0.6$ and the same values of $t$.}
\label{fig2}
\end{figure}

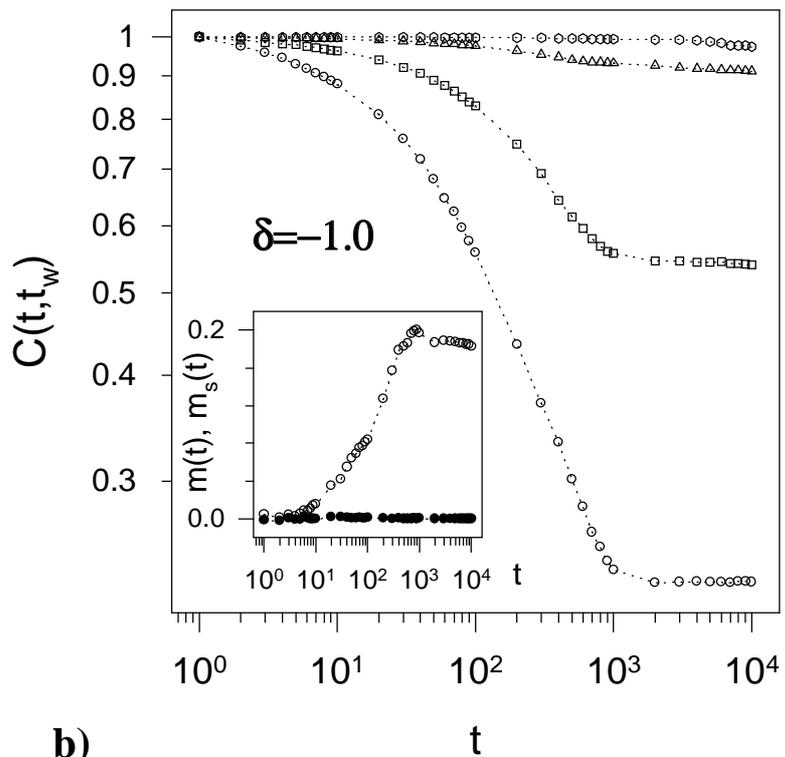
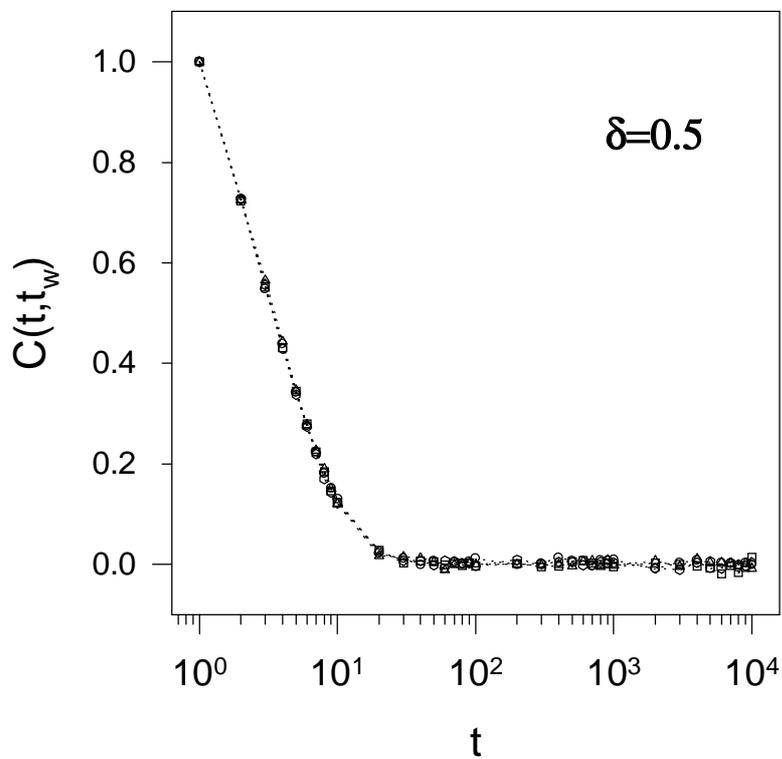
\begin{figure}
\caption{Autocorrelation function $C(t,t_w)$ {\it vs} the observation time $t$
 at $T=0.5$, for $t_w=5^2$ (circles), $t_w=5^3$ (squares), $t_w=5^4$ 
 (triangles) and $t_w=5^5$ (hexagons). (a) $\delta=-1$, inset: magnetization
 $m(t)$ (filled circles) and staggered magnetization $m_s(t)$ {\it vs} time. 
 (b) $\delta=0.5$}
\label{fig3}
\end{figure}

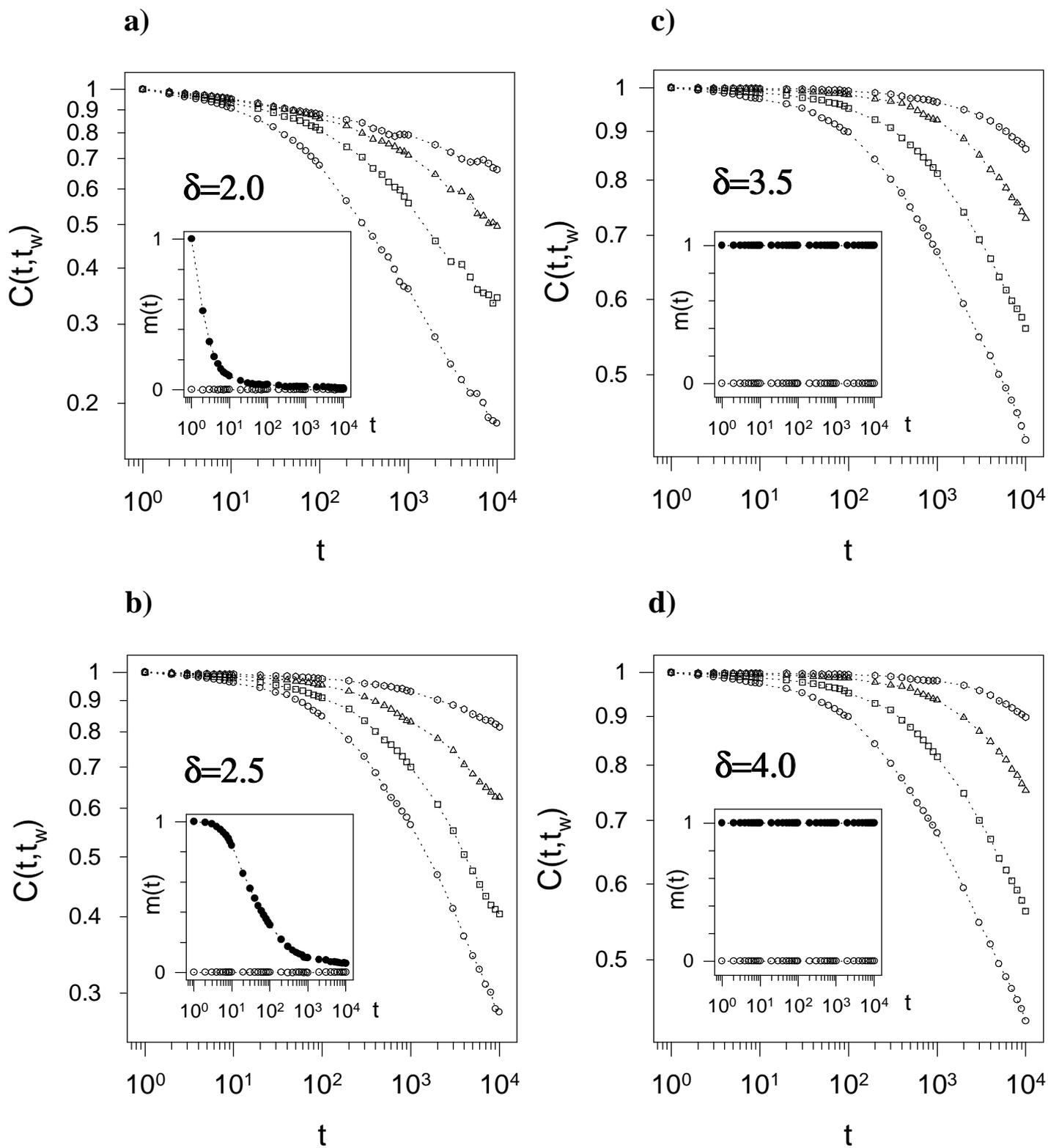
\begin{figure}
\caption{Autocorrelation function $C(t,t_w)$ {\it vs} the observation time 
 $t$ for $t_w=5^3$ (circles), $t_w=5^4$ (squares), $t_w=5^5$ (triangles) and 
 $t_w=5^6$ (hexagons), for different values of $\delta$ corresponding to the 
 striped phase. Insets: magnetization $m(t)$ {\it vs} time starting from 
 random ($m(0)=0$, open circles) and a fully magnetized ($m(0)=1$, filled 
 circles) states.}
\label{fig4}
\end{figure}

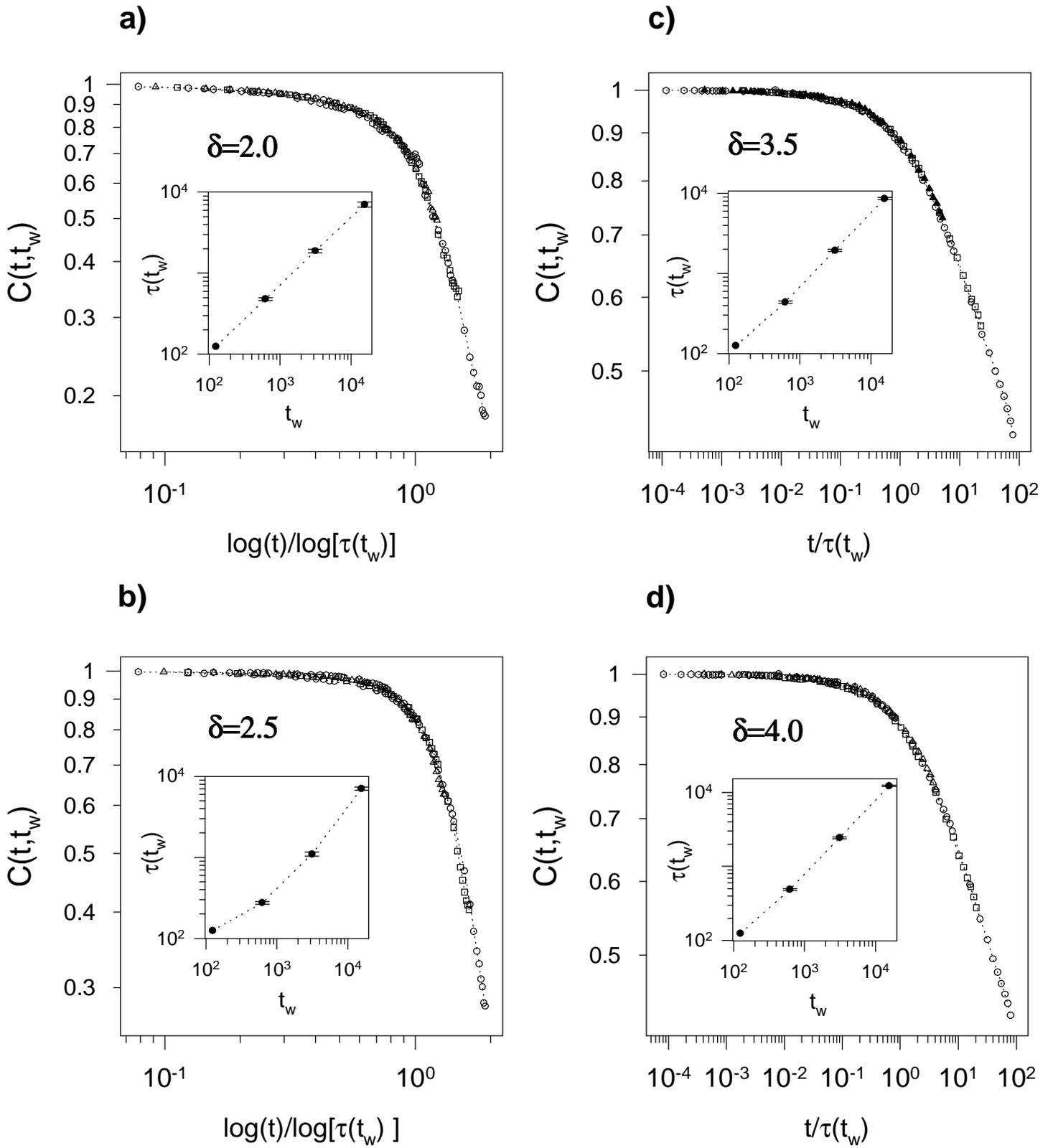
\begin{figure}
\caption{Data collapse of the curves shown in Fig.4 (the values of $t_w$ and 
 the corresponding symbols are the same in both figures).}
\label{fig5}
\end{figure}

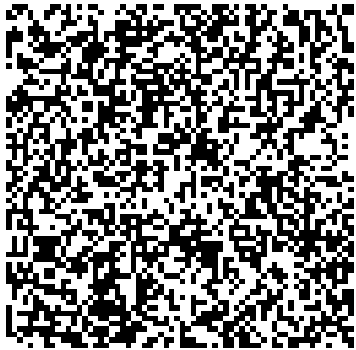
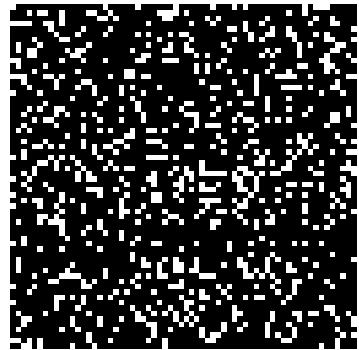
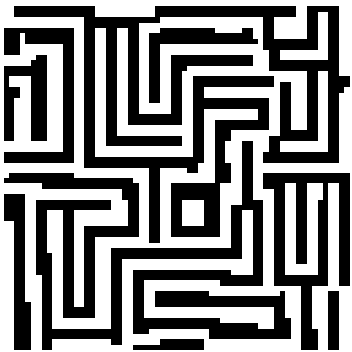
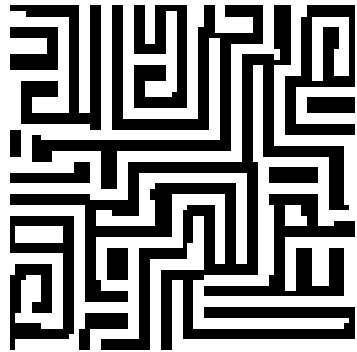
\end{document}